\documentstyle[12pt]{article}
\begin{document}
\begin{flushright}
To appear in {\it J. Phys. G}
\end{flushright}
\vspace*{0.5cm}

\begin{center}
{\bf QUARK FAMILY DISCRIMINATION AND\\
FLAVOUR-CHANGING NEUTRAL CURRENTS IN THE\\
   $\mbox{SU(3)}_C \otimes \mbox{SU(3)}_L
\otimes \mbox{U(1)}$ MODEL
WITH RIGHT-HANDED NEUTRINOS}\\
\vspace{2cm}
{\bf Hoang Ngoc Long} and {\bf Vo Thanh Van}\\
{\it Institute of  Physics,
National Centre for Natural Science and 
Technology,\\
P.O.Box 429, Bo Ho, Hanoi 10000, Vietnam}\\
\vspace{1cm}

Abstract\\
\end{center}
Contributions of flavour-changing neutral currents in 
the 3 3 1 model with
right-handed neutrinos to mass difference of the neutral 
meson system  $\Delta m_P  (P = K, D, B)$ are calculated.
Using the Fritzsch anzats on quark mixing, we show that
the third family should  be different from the first two. 
We obtain a lower bound on mass of the new heavy neutral
gauge boson as 1.02 TeV.

PACS number(s): 12.15.-y, 13.15.J\\
\vspace{2cm}

\noindent
{\large\bf I. Introduction}\\[0.3cm]
\hspace*{0.5cm}The standard model (SM) has been very successful
in explaing high energy phenomena. However there still remain
some important questions we should understand. It is especially 
necessary to answer the quark family problem and 
hierarchy puzzle.
  In addition, the SuperKamiokande atmospheric neutrino 
data~\cite{superk} provided an evidence for neutrino 
oscillation and consequently
non-zero neutrino mass. It is known that, neutrinos are 
massless in the SM, therefore the SuperKamiokande 
result calls  firstly  for the SM extension.

Among the possible extensions, the models based on the
 $\mbox{SU(3)}_C \otimes \mbox{SU(3)}_L\otimes \mbox{U(1)}_N$
(3 3 1) gauge group~\cite{ppf,flt} have some interesting 
properties such as: first, it can explain why family number 
$N$ is equal to three. Second, one quark family  is 
treated differently from the other two, and this  
gives some indication as to why the top quark is 
unbalancing heavy.
Third, the Peccei-Quinn symmetry, necessary to solve the 
strong-CP problem, follows naturally from particle 
content in these models~\cite{pal}.

There are two main versions of 3 3 1 models:
the minimal~\cite{ppf,dng} in which all lepton components 
($\nu, l, (l_L)^c$) belong to the lepton triplet and a variant,
in which right-handed neutrinos (r.h.neutrinos) are included
i.e. ($\nu, l, (\nu)_L^c$) (hereafter we call it the
model with r.h. neutrinos).

The fact that one quark family is treated 
differently from the other two,  leads to the 
flavour-changing neutral currents (FCNC's), which give a
contribution to the mass difference of the neural
meson systems at the tree level.
The effect in the minimal model was considered in~\cite{liu}.
In this paper we  shall consider the FCNC's effect  
in the model with r.h. neutrinos.

\noindent
{\large\bf II. The model }\\[0.3cm]
\hspace*{0.5cm} Let us briefly recapitulate the basic elements
of the model. Fermions are in triplet
\begin{equation}
f^{a}_L = \left( \begin{array}{c}
               \nu^a_L\\ e^a_L\\ (\nu^c_L)^a
 \end{array}  \right) \sim (1, 3, -1/3), e^a_R\sim (1, 1, -1),
\label{l}
\end{equation} 
where a = 1, 2, 3 is the family index.

Two of the three quark families transform identically and 
one family (it does not matter which one) 
transforms in a different representation of the gauge group 
 $\mbox{SU(3)}_C \otimes \mbox{SU(3)}_L\otimes 
\mbox{U(1)}_N$:
\begin{equation}
Q_{iL} = \left( \begin{array}{c}
                d_{iL}\\-u_{iL}\\ D_{iL}\\ 
                \end{array}  \right) \sim (3, \bar{3}, 0),
\label{q}
\end{equation}
\[ u_{iR}\sim (3, 1, 2/3), d_{iR}\sim (3, 1, -1/3), 
D_{iR}\sim (3, 1, -1/3),\ i=1,2,\]
\[ Q_{3L} = \left( \begin{array}{c}
                 u_{3L}\\ d_{3L}\\ T_{L}
                \end{array}  \right) \sim (3, 3, 1/3),\]
\[ u_{3R}\sim (3, 1, 2/3), d_{3R} \sim (3, 1, -1/3), 
T_{R}\sim (3, 1, 2/3),\]
where $D$ and $T$ are exotic quarks with electric
charges $-\frac{1}{3}$ and  $\frac{2}{3}$,
respectively.

Fermion mass generation and symmetry breaking can be 
achieved with just three $SU(3)_{L}$ triplets
\begin{equation}
\chi = \left( \begin{array}{c}
                \chi^o\\ \chi^-\\ \chi^{,o}\\ 
                \end{array}  \right) \sim (1, 3, -1/3),
\rho = \left( \begin{array}{c}
                \rho^+\\ \rho^o\\ \rho^{,+}\\ 
                \end{array}  \right) \sim (1, 3, 2/3),\
\eta = \left( \begin{array}{c}
                \eta^o\\ \eta^-\\ \eta^{,o}\\ 
                \end{array}  \right) \sim (1, 3, -1/3).
\label{h2}
\end{equation}
All the Yukawa terms of quarks are given
\begin{eqnarray}
{\cal L}_{Yuk}^{\chi}&=&\lambda_1\bar{Q}_{3L}T_{R}\chi +
 \lambda_{2ij}\bar{Q}_{iL}d^{'}_{jR}\chi^{*} + 
\mbox{h.c.}\nonumber\\
 &=&\lambda_1(\bar{u}_{3L}\chi^o+\bar{d}_{3L}\chi^-
+\bar{T}_L\chi^{,o})T_R
+\lambda_{2ij}(\bar{d}_{iL}\chi^{o*}-\bar{u}_{iL}\chi^++
\bar{D}_{iL}\chi^{,o*})D_{jR} + \mbox{h.c.}\nonumber\\
{\cal L}_{Yuk}^{\eta}&=&\lambda_{3a}\bar{Q}_{3L}
u_{aR}\eta+\lambda_{4ia}\bar{Q}_{iL}
d_{aR}\eta^{*}+\mbox{h.c.}\nonumber\\
&=&\lambda_{3a}(\bar{u}_{3L}\eta^o+
\bar{d}_{3L}\eta^-+\bar{T}_L\eta^{,o})
u_{aR}+\lambda_{4ia}(\bar{d}_{iL}
\eta^{o*}-\bar{u}_{iL}\eta^+
+\bar{D}_{iL}\eta^{,o*})d_{aR}+\mbox{h.c.}\nonumber\\
{\cal L}_{Yuk}^{\rho}&=&\lambda_{1a}
\bar{Q}_{3L}d_{aR}\rho +
 \lambda_{2ia}\bar{Q}_{iL}u_{aR}\rho^{*}\nonumber\\
&=&\lambda_{1a}(\bar{u}_{3L}\rho^++\bar{d}_{3L}\rho^o+
\bar{T}_L\rho^{,+})d_{aR}+
\lambda_{2ia}(\bar{D}_{iL}\rho^- 
-\bar{u}_{iL}\rho^{o*}\nonumber\\
& &+\bar{D}_{iL}\rho^{,-})u_{aR}
+\mbox{h.c.}.
\label{yukawa}
\end{eqnarray}
In this model the Higgs triplets in Eq.~(\ref{h2}) 
should develop  VEVs as follow:
\[\langle\chi \rangle^T = (0, 0, \omega/\sqrt{2}),\
\langle\rho \rangle^T = (0, u/\sqrt{2}, 0),\
\langle\eta \rangle^T = (v/\sqrt{2}, 0, 0).\]

The new complex gauge bosons in this model are 
$\sqrt{2} X^0_\mu = W^4_\mu - i W^5_\mu,
\sqrt{2} Y^+_\mu = W^6_\mu - i W^7_\mu$. Both these
bosons carry lepton number two, hence they are called 
bileptons. In~\cite{li} the first constraints 
on masses of the bileptons
in this model  are derived by considering $S, T$ parameters:
$213\  {\rm GeV} \leq M_{Y^+} \leq 234\  
{\rm GeV},\  230\  {\rm GeV}  \leq M_{X^0} \leq 251$  GeV.

The physical neutral gauge bosons are mixtures of $Z, Z'$:
\begin{eqnarray}
Z^1  &=&Z\cos\phi - Z'\sin\phi,\nonumber\\
Z^2  &=&Z\sin\phi + Z'\cos\phi,
\end{eqnarray}
where the photon field $A_\mu$ and $Z,Z'$ are given by:
\begin{eqnarray}
A_\mu  &=& s_W  W_{\mu}^3 + c_W\left(-\frac{t_W}{\sqrt{3}}
\ W^8_{\mu} +\sqrt{1-\frac{t^2_W}{3}}\  
B_{\mu}\right),\nonumber\\
Z_\mu  &=& c_W  W^3_{\mu} - s_W\left(-\frac{t_W}{\sqrt{3}}\ 
W^8_{\mu}+ \sqrt{1-\frac{t_W^2}{3}}\  B_{\mu}\right), \\
\label{apstat}
Z'_\mu &=& \sqrt{1-\frac{t_W^2}{3}}\  W^8_{\mu}+
\frac{t_W}{\sqrt{3}}\ B_{\mu}\nonumber.
\end{eqnarray}
Here $s_W$ stands for  $ \sin \theta_W$. The mixing angle 
$\phi$ is defined by
\begin{equation}
\tan^2\phi =\frac{m_{Z}^2-m^2_{Z^1}}{M_{Z^2}^2-m_{Z}^2},
\label{tphi}
\end{equation}
where $m_{Z^1}$ and $M_{Z^2}$ are the 
{\it physical} mass eigenvalues.

The interactions among fermions and  $Z_1, Z_2$  
are given as follows:
\begin{eqnarray}
{\cal L}^{NC}&=&\frac{g}{2c_W}\left\{\bar{f}\gamma^{\mu} 
[a_{1L}(f)(1-\gamma_5) + a_{1R}(f)(1+\gamma_5)]f 
Z^1_{\mu}\right.\nonumber\\
             & &+ \left.\bar{f}\gamma^{\mu} 
[a_{2L}(f)(1-\gamma_5) + a_{2R}(f)(1+\gamma_5)]f 
Z^2_{\mu}\right\}.
\label{nc}
\end{eqnarray}
where
\begin{eqnarray}
a_{1L,R}(f) &=&\cos\phi\ [T^3(f_{L,R})-s_W^2 Q(f)]\nonumber\\
           & &- c_W^2\left[\frac{3N(f_{L,R})}{(3-4s_W^2)^{1/2}}
-\frac{(3-4s_W^2)^{1/2}}{2c^2_W}Y(f_{L,R})\right]
\sin\phi,\nonumber\\
a_{2L,R}(f)&=& c_W^2\left[\frac{3N(f_{L,R})}{(3-4s_W^2)^{1/2}}
-\frac{(3-4s_W^2)^{1/2}}{2c^2_W}Y(f_{L,R})\right]
\cos\phi\nonumber\\
           & &+ \sin\phi\ [T^3(f_{L,R})-s_W^2 Q(f)].
\label{vaz}
\end{eqnarray}
Here $T^3(f)$ and $Q(f)$ are, respectively, the third 
component of the weak isospin and the 
charge of the fermion $f$. 

\noindent
{\large\bf III. Flavour-changing neutral currents and 
mass difference of the neutral meson systems}\\[0.3cm]
\hspace*{0.5cm}
Due to the fact that one family of left-handed quarks is treated 
differently from the other two, the N charges for 
left-handed quarks are different too (see Eq.~(\ref{q})). 
Therefore flavour-changing neutral currents $Z_1, Z_2$ 
occur through a mismatch between weak and mass eigenstates.

Let us diagolize mass matrices by three biunitary 
transformations  
\begin{eqnarray}
U'_L & = & V_L^U U_L,\   U'_R = V_R^U U_R,\nonumber\\
D'_L & = & V_L^D D_L,\   D'_R = V_R^D D_R,
\label{tran}
\end{eqnarray}
where $U \equiv (u, c, t)^T, \ D \equiv (d,s,b)^T$.\\
The usual Cabibbo-Kobayashi-Maskawa matrix is given by
\begin{equation}
V_{CKM} = V_L^{U+} V_L^D.
\label{vckm}
\end{equation}

Using unitarity of the $V^D$ and $V^U$ matrices, we get 
flavour-changing neutral interactions
\begin{eqnarray}
{\cal L}^{NC}_{ds}&=&\frac{g c_W}{2 \sqrt{3-4 s_W^2}}
\left[V^{D*}_{Lid} 
V^D_{Lis}\right] \bar{d}_L \gamma_\mu s_L 
\left(\cos \phi Z_2^\mu - \sin \phi Z_1^\mu \right ),
\nonumber\\
{\cal L}^{NC}_{uc}&=&\frac{g c_W}{2 \sqrt{3-4 s_W^2}}
\left[V^{U*}_{Liu} V^U_{Lic}\right] \bar{u}_L \gamma_\mu c_L 
\left(\cos \phi Z_2^\mu - \sin \phi Z_1^\mu \right ),\\
{\cal L}^{NC}_{db}&=&\frac{g c_W}{2 \sqrt{3-4 s_W^2}}
\left[V^{D*}_{Lid} V^D_{Lib}\right] \bar{d}_L \gamma_\mu b_L 
\left(\cos \phi Z_2^\mu - \sin \phi Z_1^\mu \right )\nonumber,
\label{fcnc}
\end{eqnarray}
where $i$ denotes the number of "different'' quark family
i.e. the $ \mbox{SU(3)}_L $ quark triplet.  

For the neutral kaon system, we get then  
effective Lagrangian  
\begin{equation}
{\cal L}^{\Delta S=2}_{eff} = \frac{\sqrt{2} G_F\  c_W^4 
\cos^2 \phi}{(3-4 s_W^2)}\left[V^{D*}_{Lid} 
V^D_{Lis}\right]^2| \bar{d}_L \gamma^\mu s_L|^2
\left( \frac{m^2_{Z_1}}{M^2_{Z_2}} + \tan^2 \phi \right).
\label{eff}
\end{equation}
Similar expressions can be easily written out for $D^0 - 
\bar{D}^0$ and  $B^0 - \bar{B}^0$ systems.
From ~(\ref{eff}) it is straightforward 
to get the mass difference
\begin{equation}
\Delta m_P =  \frac{4 G_F\   c_W^4\  
\cos^2 \phi}{3 \sqrt{2} 
 (3-4 s_W^2)}\left[V^{D*}_{Lid} 
V^D_{Li\alpha}\right]^2 \left( 
\frac{m^2_{Z_1}}{M^2_{Z_2}} + 
 \tan^2 \phi \right) f^2_P B_P m_P,
\label{masdif}
\end{equation}
where $\alpha = s$ for the  $K_L - K_S$ and $\alpha = b$ 
for  the $B^0 - \bar{B}^0$
mixing systems. The $D^0 - \bar{D}^0$ mass difference 
is given by the expression for  the $K^0$ system  
with replace of $V^D$ by $V^U$.
The $Z - Z'$ mixing angle $\phi$ was bounded
and to be~\cite{flt,dng}: $ |\phi| \leq 10^{-3} $,
hence if $M_{Z_2}$ is in order of one hundred TeV,
the $Z - Z'$ mixing has to be taken into account. 

In the usual case, the $Z-Z'$ mixing is constrained to 
be very small, it can be safely neglected. 
Therefore FCNC's occur only via $Z_2$ couplings.
For the shothand hereafter we rename  $Z_1$ to be $Z$
and  $Z_2$ to be $Z'$.

Since it is generally recognized that the most stringent limit
from $\Delta m_K$, we shall mainly discuss this quantity.
We  use the experimental values~\cite{caso}
\begin{eqnarray}
\Delta m_K & = & ( 3.489 \pm 0.009)\times 10^{- 12}\  
{\rm MeV}, \hspace*{1cm}  m_K \simeq 498\   {\rm MeV} 
\label{data}
\end{eqnarray}
and 
\begin{eqnarray}
\sqrt{B_K} f_K & = & 135 \pm 19 \ {\rm MeV}.
\label{fb}
\end{eqnarray}

Following the idea of Gaillard and Lee~\cite{gali}, 
it is reasonable to expect that $Z'$ exchange 
contributes a $\Delta m$ no larger
than observed values. Substituting~(\ref{data}) and~(\ref{fb})
 into~(\ref{masdif}) we get
\begin{eqnarray}
M_{Z'}& >& 2.63 \times 10^5\  \eta_{Z'}  \left[Re(V^{D*}_{Lid} 
V^D_{Lis})^2\right]^{1/2} \  {\rm GeV}.
\label{gh}
\end{eqnarray}
where $\eta_{Z'} \approx  0.55$ is
the leading order QCD corrections~\cite{gwi}.

Let us call $ \Delta m_K^{min},  \Delta m_K^{rhn}$ 
contributions to $\Delta m$  from  the $Z'$ in the  minimal
3 3 1 model and in the model  with r.h. neutrinos, 
respectively. We have then 
\begin{equation}
R \equiv \frac{ \Delta m_K^{min}}{ \Delta m_K^{rhn}}
= \frac{2 (3 - 4 s_W^2) }{3 ( 1 - 4 s^2_W)} = 19.7,
\label{rel}
\end{equation}
for~\cite{caso} $s_W^2 = 0.2312$.  Because of the 
denominator, the relation is highly sensitive to the value of 
the Weinberg angle. It is easy to see that a 
limit for the $Z'$ following from Eq~(\ref{gh}) 
in the model with r.h.neutrinos is approximately 4.4 
times  smaller than that in the minimal version.

From the present experimental data we cannot get
the constraint on $V^{U,D}_{Lij}$. These
matrix elemetns are only constrainted by ~(\ref{vckm}).
However, it would seem more natural, if Higgs scalars are
associated with fermion generations, to have the choice of 
nondiagonal elements depends on the fields to which 
the Higgs scalars couple. By this way, the 
simple Fritzsch~\cite{hf} scheme gives us
\begin{equation}
V^D_{ij} \approx \left( \frac{m_i}{m_j} \right)^{1/2},
\hspace*{1cm} i < j.
\label{hfr}
\end{equation}
 
Combining ~(\ref{gh}) and ~(\ref{hfr}) we get 
the following bounds on $M_{Z'}$:
\begin{eqnarray}
M_{Z'}& \geq & 38\   {\rm  TeV}, \hspace*{0.3cm} {\rm if\ the\
first\ or\ the \ second\ quark\ family\ is\ 
different\ (\ in\ triplet)}
\nonumber\\
M_{Z'}& \geq& 1.02 \  {\rm  TeV}, \hspace*{0.3cm} 
{\rm if\ the\ third\ quark\ family\ is\ different}
\label{thu}
\end{eqnarray}
From  ~(\ref{thu}) we see that  to keep relatively 
low bounds on  $M_{Z'}$ the third family 
should be the one that is different
from the other two i.e. is in triplet.

Our numerical estimation is based on the fact that 
all the phases of the matrix elements equal to zero. 
The inclusion of complex phases would induce to a 
reduction in the  mass limit. However
the hierarchical picture should not be modified.

\noindent
{\large\bf IV. Summary}\\[0.3cm]
\hspace*{0.5cm} We have studied the FCNC's
 in the 3 3 1 model with r.h. neutrinos arisen from 
the family discrimination in this model. This gives a 
reason to conclude that the third family should 
be treated differently from the first two. 
In this sense, the $\Delta m_K$
gives us the lower bound on $M_{Z'}$ as 1.02 TeV.
It is to be mentioned that our conclusion is similar
with that in the minimal version~\cite{dpp}:
\begin{eqnarray}
M_{Z'}& \geq & 315\   {\rm  TeV}, \hspace*{0.3cm} {\rm if\ the\
first\ or\ the \ second\ quark\ family\ is\ 
different\ (\ in\ triplet)}
\nonumber\\
M_{Z'}& \geq& 10 \  {\rm  TeV}, \hspace*{0.3cm} 
{\rm if\ the\ third\ quark\ family\ is\ different}
\nonumber
\end{eqnarray}

It is interesting to note that in the Fritzsch anzats, 
the limits for $M_{Z'}$ following from $\Delta m_B$ 
are independent of the family choice.

We emphasize that the FCNC's in the minimal model are larger
than those in the considered version due to the factor
$\frac{1}{\left(1 - 4 s_W^2\right)}$. Therefore the 
lower bounds on  $M_{Z'}$ are smaller accordingly.

In both versions of the 3 3 1 models, the third quark family 
should be different from the first two, and this gives us some 
indication of why the top quark is so heavy.

{\bf Acknowledgements}

This work has been  initialed when the first author (H.N.L) 
was at the Department of Physics, Chuo University, Tokyo.
He expresses sincere gratitude to Professor T. Inami for 
helpful discussions and  warm hospitality.
This work was supported in part by Research Programme under grant
$N^0$ : QT 98.04 and KT - 04.1.2.

\end{document}